\documentclass{Interspeech}

\interspeechcameraready

\title{You Are What You Say: Exploiting Linguistic Content for\\VoicePrivacy Attacks}

\author[affiliation={1}]{Ünal Ege}{Gaznepoglu}
\author[affiliation={2}]{Anna}{Leschanowsky}
\author[affiliation={3}]{Ahmad}{Aloradi}
\author[affiliation={2}]{Prachi}{Singh}
\author[affiliation={3}]{Daniel}{Tenbrinck}
\author[affiliation={1}]{Emanuël A. P.}{Habets}
\author[affiliation={4}]{Nils}{Peters}

\affiliation{International Audio Laboratories Erlangen}{FAU Erlangen-Nürnberg}{Germany}
\affiliation{}{Fraunhofer Institute for Integrated Circuits (IIS), Erlangen}{Germany}
\affiliation{Department of Data Science}{FAU Erlangen-Nürnberg}{Germany}
\affiliation{Department of Electrical and Electronics Engineering}{Trinity College Dublin}{Ireland}
\email{ege.gaznepoglu@audiolabs-erlangen.de}

\usepackage{acronym}
\usepackage{easyReview}
\usepackage{subfig}
\usepackage[style=ieee, doi=false, isbn=false, url=true, eprint=false, related=false, maxnames=6, minnames=1, date=year, block=space]{biblatex}
\usepackage{pbalance}
\AtEveryBibitem{%
    \ifboolexpr{
    test {\ifentrytype{misc}}
    }
    {}
    {\clearfield{url}}
}
\AtEveryBibitem{\clearfield{urlyear}}
\AtEveryBibitem{\clearlist{language}}
\AtEveryBibitem{\clearfield{note}}
\AtEveryBibitem{\clearlist{location}}
\AtEveryBibitem{\clearfield{location}}
\AtEveryBibitem{\clearfield{publisher}}
\AtEveryBibitem{\clearlist{publisher}}
\AtEveryBibitem{\clearlist{pages}}
\AtEveryBibitem{\clearfield{pages}}
\AtEveryBibitem{\clearfield{pages}}
\usepackage[hang,flushmargin]{footmisc}

\usetikzlibrary{arrows.meta}

\makeatletter

\pgfdeclarearrow{
  name = ipe _linear,
  defaults = {
    length = +1bp,
    width  = +.666bp,
    line width = +0pt 1,
  },
  setup code = {
    \pgfarrowssetbackend{0pt}
    \pgfarrowssetvisualbackend{
      \pgfarrowlength\advance\pgf@x by-.5\pgfarrowlinewidth}
    \pgfarrowssetlineend{\pgfarrowlength}
    \ifpgfarrowreversed
      \pgfarrowssetlineend{\pgfarrowlength\advance\pgf@x by-.5\pgfarrowlinewidth}
    \fi
    \pgfarrowssettipend{\pgfarrowlength}
    \pgfarrowshullpoint{\pgfarrowlength}{0pt}
    \pgfarrowsupperhullpoint{0pt}{.5\pgfarrowwidth}
    \pgfarrowssavethe\pgfarrowlinewidth
    \pgfarrowssavethe\pgfarrowlength
    \pgfarrowssavethe\pgfarrowwidth
  },
  drawing code = {
    \pgfsetdash{}{+0pt}
    \ifdim\pgfarrowlinewidth=\pgflinewidth\else\pgfsetlinewidth{+\pgfarrowlinewidth}\fi
    \pgfpathmoveto{\pgfqpoint{0pt}{.5\pgfarrowwidth}}
    \pgfpathlineto{\pgfqpoint{\pgfarrowlength}{0pt}}
    \pgfpathlineto{\pgfqpoint{0pt}{-.5\pgfarrowwidth}}
    \pgfusepathqstroke
  },
  parameters = {
    \the\pgfarrowlinewidth,%
    \the\pgfarrowlength,%
    \the\pgfarrowwidth,%
  },
}

\pgfdeclarearrow{
  name = ipe _pointed,
  defaults = {
    length = +1bp,
    width  = +.666bp,
    inset  = +.2bp,
    line width = +0pt 1,
  },
  setup code = {
    \pgfarrowssetbackend{0pt}
    \pgfarrowssetvisualbackend{\pgfarrowinset}
    \pgfarrowssetlineend{\pgfarrowinset}
    \ifpgfarrowreversed
      \pgfarrowssetlineend{\pgfarrowlength}
    \fi
    \pgfarrowssettipend{\pgfarrowlength}
    \pgfarrowshullpoint{\pgfarrowlength}{0pt}
    \pgfarrowsupperhullpoint{0pt}{.5\pgfarrowwidth}
    \pgfarrowshullpoint{\pgfarrowinset}{0pt}
    \pgfarrowssavethe\pgfarrowinset
    \pgfarrowssavethe\pgfarrowlinewidth
    \pgfarrowssavethe\pgfarrowlength
    \pgfarrowssavethe\pgfarrowwidth
  },
  drawing code = {
    \pgfsetdash{}{+0pt}
    \ifdim\pgfarrowlinewidth=\pgflinewidth\else\pgfsetlinewidth{+\pgfarrowlinewidth}\fi
    \pgfpathmoveto{\pgfqpoint{\pgfarrowlength}{0pt}}
    \pgfpathlineto{\pgfqpoint{0pt}{.5\pgfarrowwidth}}
    \pgfpathlineto{\pgfqpoint{\pgfarrowinset}{0pt}}
    \pgfpathlineto{\pgfqpoint{0pt}{-.5\pgfarrowwidth}}
    \pgfpathclose
    \ifpgfarrowopen
      \pgfusepathqstroke
    \else
      \ifdim\pgfarrowlinewidth>0pt\pgfusepathqfillstroke\else\pgfusepathqfill\fi
    \fi
  },
  parameters = {
    \the\pgfarrowlinewidth,%
    \the\pgfarrowlength,%
    \the\pgfarrowwidth,%
    \the\pgfarrowinset,%
    \ifpgfarrowopen o\fi%
  },
}

\newdimen\ipeminipagewidth

\tikzstyle{ipe import} = [
  x=1bp, y=1bp,
  ipe node stretch/.store in=\ipenodestretch,
  ipe stretch normal/.style={ipe node stretch=1},
  ipe stretch normal,
  ipe node/.style={
    anchor=base west, inner sep=0, outer sep=0, scale=\ipenodestretch
  },
  ipe mark scale/.store in=\ipemarkscale,
  ipe mark tiny/.style={ipe mark scale=1.1},
  ipe mark small/.style={ipe mark scale=2},
  ipe mark normal/.style={ipe mark scale=3},
  ipe mark large/.style={ipe mark scale=5},
  ipe mark normal, 
  ipe circle/.pic={
    \draw[line width=0.2*\ipemarkscale]
      (0,0) circle[radius=0.5*\ipemarkscale];
    \coordinate () at (0,0);
  },
  ipe disk/.pic={
    \fill (0,0) circle[radius=0.6*\ipemarkscale];
    \coordinate () at (0,0);
  },
  ipe fdisk/.pic={
    \filldraw[line width=0.2*\ipemarkscale]
      (0,0) circle[radius=0.5*\ipemarkscale];
    \coordinate () at (0,0);
  },
  ipe box/.pic={
    \draw[line width=0.2*\ipemarkscale, line join=miter]
      (-.5*\ipemarkscale,-.5*\ipemarkscale) rectangle
      ( .5*\ipemarkscale, .5*\ipemarkscale);
    \coordinate () at (0,0);
  },
  ipe square/.pic={
    \fill
      (-.6*\ipemarkscale,-.6*\ipemarkscale) rectangle
      ( .6*\ipemarkscale, .6*\ipemarkscale);
    \coordinate () at (0,0);
  },
  ipe fsquare/.pic={
    \filldraw[line width=0.2*\ipemarkscale, line join=miter]
      (-.5*\ipemarkscale,-.5*\ipemarkscale) rectangle
      ( .5*\ipemarkscale, .5*\ipemarkscale);
    \coordinate () at (0,0);
  },
  ipe cross/.pic={
    \draw[line width=0.2*\ipemarkscale, line cap=butt]
      (-.5*\ipemarkscale,-.5*\ipemarkscale) --
      ( .5*\ipemarkscale, .5*\ipemarkscale)
      (-.5*\ipemarkscale, .5*\ipemarkscale) --
      ( .5*\ipemarkscale,-.5*\ipemarkscale);
    \coordinate () at (0,0);
  },
  /pgf/arrow keys/.cd,
  ipe arrow normal/.style={scale=1},
  ipe arrow tiny/.style={scale=.4},
  ipe arrow small/.style={scale=.7},
  ipe arrow large/.style={scale=1.4},
  ipe arrow normal,
  /tikz/.cd,
  ipe arrows/.style={
    ipe normal/.tip={
      ipe _pointed[length=1bp, width=.666bp, inset=0bp,
                   quick, ipe arrow normal]},
    ipe pointed/.tip={
      ipe _pointed[length=1bp, width=.666bp, inset=0.2bp,
                   quick, ipe arrow normal]},
    ipe linear/.tip={
      ipe _linear[length = 1bp, width=.666bp,
                  ipe arrow normal, quick]},
    ipe fnormal/.tip={ipe normal[fill=white]},
    ipe fpointed/.tip={ipe pointed[fill=white]},
    ipe double/.tip={ipe normal[] ipe normal},
    ipe fdouble/.tip={ipe fnormal[] ipe fnormal},
    ipe arc/.tip={ipe normal},
    ipe farc/.tip={ipe fnormal},
    ipe ptarc/.tip={ipe pointed},
    ipe fptarc/.tip={ipe fpointed},
  },
  ipe arrows, 
]

\tikzset{
  rgb color/.code args={#1=#2}{%
    \definecolor{tempcolor-#1}{rgb}{#2}%
    \tikzset{#1=tempcolor-#1}%
  },
}

\makeatother

\usetikzlibrary{arrows.meta,patterns}
\usetikzlibrary{ipe} 

\everymath=\expandafter{\the\everymath\displaystyle}
\makeatletter\@ifpackageloaded{underscore}{}{\usepackage[strings]{underscore}}\makeatother

\begin{document}
\newacro{BN}{bottleneck feature}
\newacro{RTF}{real time factor}
\newacro{DNN}{deep neural network}
\newacro{F0}{fundamental frequency}
\newacro{EER}{equal error rate}
\newacro{WER}{word error rate}
\newacro{PI}{personal information}
\newacro{MSE}{mean-squared error}
\newacro{NSF}{neural source-filter}
\newacro{AM}{acoustic model}
\newacro{VPC}{Voice Privacy Challenge}
\newacro{ASV}{automatic speaker verification}
\newacro{ASR}{automatic speech recognition}
\newacro{TDNN}{time delay neural network}
\newacro{FPE}{Fine Pitch Error}
\newacro{GPE}{Gross Pitch Error}
\newacro{MFCC}{Mel Frequency Cepstral Coefficients}
\newacro{RTF}{Real Time Framework}
\newacro{NSF}{Neural Source-Filter}
\newacro{VC}{voice conversion}
\newacro{AE}{autoencoder}
\newacro{CNN}{convolutional neural network}
\newacro{PPG}{phonetic posteriorgrams}
\newacro{UAR}{unweighted average recall}
\newacro{PQMF}{pseudo quadrature mirror filterbank}
\newacro{OOD}{out-of-distribution}
\newacro{SER}{speech emotion recognition}
\newacro{OHNN}{orthogonal Householder neural network}
\newacro{GAN}{generative adversarial network}
\newacro{AAM}{additive angular margin}
\newacro{ZEBRA}{Zero Evidence Biometric Recognition Assessment}
\maketitle

%
\begin{abstract} 
    Speaker anonymization systems hide the identity of speakers while preserving other information such as linguistic content and emotions. To evaluate their privacy benefits, attacks in the form of \ac{ASV} systems are employed. In this study, we assess the impact of intra-speaker linguistic content similarity in the attacker training and evaluation datasets, by adapting BERT, a language model, as an ASV system. On the VoicePrivacy Attacker Challenge datasets, our method achieves a mean \ac{EER} of 35\%, with certain speakers attaining \acp{EER} as low as 2\%, based solely on the textual content of their utterances. Our explainability study reveals that the system decisions are linked to semantically similar keywords within utterances, stemming from how LibriSpeech is curated. Our study suggests reworking the VoicePrivacy datasets to ensure a fair and unbiased evaluation and challenge the reliance on global \ac{EER} for privacy evaluations.
\end{abstract}
\acresetall

\section{Introduction}

The field of speaker anonymization has emerged in response to the risks associated with advances in speech-processing technology, such as the inadvertent disclosure of personal information (e.g., age, health) when using cloud-enabled voice interfaces \cite{tomashenko_introducing_2020}. Speaker anonymization systems protect the speaker's identity while preserving important information for downstream tasks such as \ac{ASR} and emotion recognition \cite{champion_vpc_evalplan_2024}. Recently, a VoicePrivacy Attacker Challenge \cite{tomashenko_vpac_evalplan_2025} was held for the first time. Its aim is to develop techniques to compromise the privacy of speakers that have been processed by seven anonymization systems. These include the top three baseline systems from the \ac{VPC} 2024: B3, B4, B5, and the top four participant-submitted systems.

Works on both anonymization \cite{meyer_prosody_2023, panariello_speaker_2024, champion_anonymizing_2024} and attacks to anonymization systems \cite{champion_invertibility_2021, champion_evaluating_2021, zhang_attacking_2025} use the publicly provided \ac{VPC} datasets and evaluation protocols. These protocols consist of attacks based on \ac{ASV} systems for privacy evaluation. As shown in Fig.~\ref{fig:vpc_attack_models}, in \textit{ignorant} and \textit{lazy-informed} attack models, a pretrained $\text{ASV}_\textrm{eval}$ is used, while the \textit{semi-informed} attack model uses an $\text{ASV}_\textrm{eval}^\textrm{anon}$ trained on anonymized data. The \textit{unprotected} case serves as a reference. $\text{ASV}_\textrm{eval}$ and $\text{ASV}_\textrm{eval}^\textrm{anon}$ are based on ECAPA-TDNN \cite{desplanques_ecapa-tdnn_2020}, and more information can be found in the challenge evaluation plan \cite{tomashenko_vpac_evalplan_2025}. The speaker embeddings extracted by these systems are used for enrollment and trial on corresponding \texttt{libri-dev} and \texttt{libri-test} utterances, using cosine similarity as a similarity measure. Finally, \ac{EER} is calculated for male and female speakers. Lower \acp{EER} correspond to a better de-anonymization and hence a successful attack.

\begin{figure}[!t]
    \centering
    \subfloat[Unprotected case]{\resizebox{\linewidth}{!}{\begingroup
\renewcommand{\baselinestretch}{1} \endlinechar=-1 \input{Figures/eval_unprotected.tex}\endgroup \renewcommand{\baselinestretch}{1.5}}}\\ \vspace{-0.75em}
    \subfloat[Ignorant attack model]{\resizebox{\linewidth}{!}{\begingroup
\renewcommand{\baselinestretch}{1} \endlinechar=-1 \input{Figures/eval_ignorant.tex}\endgroup \renewcommand{\baselinestretch}{1.5}}}\\ \vspace{-0.75em}
    \subfloat[Lazy-informed attack model]{\resizebox{\linewidth}{!}{\begingroup
\renewcommand{\baselinestretch}{1} \endlinechar=-1 \input{Figures/eval_lazy.tex}\endgroup \renewcommand{\baselinestretch}{1.5}}}\\ \vspace{-0.75em}
    \subfloat[Semi-informed attack model]{\resizebox{\linewidth}{!}{\begingroup
\renewcommand{\baselinestretch}{1} \endlinechar=-1 \input{Figures/eval_semi.tex}\endgroup \renewcommand{\baselinestretch}{1.5}}} \vspace{-0.75em}
    \caption{\ac{VPC} attack models define how much information is available to an attacker \cite{lal_srivastava_evaluating_2020}.} \vspace{-1em}
    \label{fig:vpc_attack_models}
\end{figure}

The literature on the analysis of attacker \ac{ASV} scores is rather limited. The \ac{ZEBRA} framework provides worst-case privacy disclosure for an individual per anonymization system but no disaggregated scores on a speaker-level \cite{nautsch_privacy_2020, noe_representing_2023}. In \cite{williams_anonymizing_2024}, Williams et al. explored speaker-level distributions of \ac{ASV} scores obtained through ignorant and lazy-informed attack models by identifying subpopulations with distinct behaviors, using the methodology first introduced in \cite{doddington_sheep_1998}. In \cite{tayebi_arasteh_addressing_2024} the authors used an ignorant attack model on anonymized pathological speech while evaluating the benefit of speaker anonymization for vulnerable subgroups. They found the standard deviation of \acp{EER} across healthy speakers to be very low, e.g., $32.26\% \pm 0.31$, for the DSP-based anonymization system \cite{patino_speaker_2021}. 

To the best of our knowledge, this study is the first to analyze semi-informed attacker scores on a speaker level to address the following questions: How does speaker-level performance vary for semi-informed attacks applied to various systems, and are there any identifiable patterns in the \ac{ASV} score distributions that can be exploited? 

\begin{figure*}[!t]
    \centering
    \resizebox{\linewidth}{!}{\begingroup
\renewcommand{\baselinestretch}{1} \endlinechar=-1 \input{Figures/vpc_scores_breakdown_interspeech.pgf}\endgroup \renewcommand{\baselinestretch}{1.5}}
    \caption{Speaker-level breakdown of $\text{ASV}_\textrm{eval}^\textrm{anon}$ (ECAPA-TDNN) scores on the \texttt{libri-dev} dataset. The top row shows female speakers, while the bottom row shows male speakers. Columns correspond to the attacked anonymization system. Bar plots denote the cosine similarity score distributions, where light green bars indicate the positive pairs (enrollment and trial speakers matching) and orange bars indicate the negative pairs (different enrollment and trial speakers). Red lines denote the threshold where the difference between Type 1 and Type 2 errors are minimal for each speaker, and the corresponding \acp{EER} (in \%) are shown by the red text.}
    \label{fig:vpc_scores_breakdown}
    \vspace{-1em}
\end{figure*}

\section{Semi-informed attack: status quo} \label{sec:status_quo}

This section presents an analysis of the $\text{ASV}_\textrm{eval}^\textrm{anon}$ scores on the speaker level. The anonymization systems to be attacked are chosen such that their architectures and intermediate representations are diverse. B3 performs any-to-any voice conversion via an ASR-TTS pipeline, where a Wasserstein GAN generates a target pseudo-identity \cite{meyer_prosody_2023}. In contrast, B4 \cite{panariello_speaker_2024} and B5 \cite{champion_anonymizing_2024} perform any-to-few voice conversion to real speakers. We have run the \ac{VPC} 2024 codebase\footnote{\url{https://github.com/Voice-Privacy-Challenge/Voice-Privacy-Challenge-2024/}} without modification to obtain anonymized utterances and the corresponding $\text{ASV}_\textrm{eval}^\textrm{anon}$ scores.

The scores, as well as the speaker-level thresholds and the corresponding \acp{EER} are visualized in Fig.~\ref{fig:vpc_scores_breakdown} for several variables. We observe a notable variability in the \acp{EER} across different speakers for each system, some going as low as 2\% yet the mean \acp{EER} are between 20\% and 35\%. Although for perfect anonymization a 50\% \ac{EER} is sufficient \cite{panariello_voiceprivacy_2024}, some speaker-system pairs attain \acp{EER} greater than 50\%, e.g., speaker \texttt{84} exhibits 62\% EER when anonymized with B4 and 73\% when anonymized with B5. Higher than necessary \acp{EER} attained by some speakers may obfuscate others with low \acp{EER} when averaged, giving a false sense of privacy. Therefore, when computing mean \acp{EER}, we propose clipping speaker \acp{EER} exceeding 50\% such that they fall into the interval $[0\%, 50\%]$, using
\begin{equation} \label{eq:clip}
    f(x) = \min(50, x).
\end{equation}

Closer inspection of Fig.~\ref{fig:vpc_scores_breakdown} reveals an odd behavior: none of the considered systems can effectively anonymize some speakers (\texttt{1673} and \texttt{652}), shown by the \ac{EER} values less than 20\%. This is quite counter-intuitive, given that the considered anonymization systems have very different architectures.  What could have caused these speakers to be de-anonymized by $\text{ASV}_\textrm{eval}^\textrm{anon}$ in all considered cases?

\section{Proposed text-based attack}
Consistent de-anonymization of specific speakers suggests the existence of persistent features that are invariant to different anonymization strategies. Upon manual inspection, we found that the texts read by speakers \texttt{1673} and \texttt{652} were on specific and unique topics, and some words recurring across their utterances. So, $\text{ASV}_\textrm{eval}^\textrm{anon}$ could be exploiting the intra-speaker linguistic content similarity in LibriSpeech. If true, this would confound the evaluations, because the anonymization systems are expected to preserve the linguistic content.

\begin{table}[!t]
    \centering
    \caption{Training hyperparameters}
    \begin{tabular}{ll}
        \toprule
        \textbf{Hyperparameter} & \textbf{Value}  \\ 
        \midrule
           Num. epochs & 20 \\
           Batch size & 256 \\
           Optimizer & \texttt{AdamW}, \texttt{lr}: \num{1e-4}\\
           \texttt{lr} scheduler & \texttt{LinearWithWarmup} \\
           Train-validation split & $90\%, 10\%$ \\
           \acs{AAM} parameters & Margin: $0.2$, Scale: $30$ \\
           Dropout probability & 0.1 \\
         \bottomrule
    \end{tabular}
    \vspace{-1em}
    \label{tab:hparams}
\end{table}

To test the feasibility of this hypothesis, we built a novel system that imitates $\text{ASV}_\textrm{eval}^\textrm{anon}$, but operating only on textual content. We use this 'text-based attack' on the ground truth transcriptions, i.e.,
 the information that an ideal anonymization system would preserve. Our attack is based on the HuggingFace implementation \cite{wolf_transformers_2020} of $\text{BERT}_{\text{BASE}}$, first introduced by \cite{devlin_bert_2019}. Fig.~\ref{fig:text_based_attack} outlines the training, enrollment, and trial phases, which are designed to be as similar as possible to $\text{ASV}_\textrm{eval}^\textrm{anon}$, and to align with the \ac{VPC} evaluation protocol. In particular, the balance between the batch size, the embedding dimensionality and the loss hyperparameters are crucial. We utilize the class \texttt{BertForSe\-quence\-Clas\-sification}, which comprises the BERT architecture and a pooling layer selecting the output token corresponding to the [CLS] input token. Then, a linear layer (called Penultimate) reduces the 768-dimensional hidden representation to 192 dimensions to ensure \ac{AAM} interacts with the embeddings in a similar fashion to $\text{ASV}_\textrm{eval}^\textrm{anon}$. Then, the classifier, a linear layer with L2-normalized weights and no bias, computes the logits using the embeddings after L2 normalizing them. Pre-existing dropout layers are kept, but the newly added layers do not use any dropout.

\begin{figure}[!t]
    \centering
    \subfloat[Training]{\resizebox{0.95\linewidth}{!}{\begingroup
\renewcommand{\baselinestretch}{1} \endlinechar=-1 \tikzstyle{ipe stylesheet} = [
  ipe import,
  even odd rule,
  line join=round,
  line cap=butt,
  ipe pen normal/.style={line width=0.4},
  ipe pen heavier/.style={line width=0.8},
  ipe pen fat/.style={line width=1.2},
  ipe pen ultrafat/.style={line width=2},
  ipe pen normal,
  ipe mark normal/.style={ipe mark scale=3},
  ipe mark large/.style={ipe mark scale=5},
  ipe mark small/.style={ipe mark scale=2},
  ipe mark tiny/.style={ipe mark scale=1.1},
  ipe mark normal,
  /pgf/arrow keys/.cd,
  ipe arrow normal/.style={scale=7},
  ipe arrow large/.style={scale=10},
  ipe arrow small/.style={scale=5},
  ipe arrow tiny/.style={scale=3},
  ipe arrow normal,
  /tikz/.cd,
  ipe arrows, 
  <->/.tip = ipe normal,
  ipe dash normal/.style={dash pattern=},
  ipe dash dotted/.style={dash pattern=on 1bp off 3bp},
  ipe dash dashed/.style={dash pattern=on 4bp off 4bp},
  ipe dash dash dotted/.style={dash pattern=on 4bp off 2bp on 1bp off 2bp},
  ipe dash dash dot dotted/.style={dash pattern=on 4bp off 2bp on 1bp off 2bp on 1bp off 2bp},
  ipe dash normal,
  ipe node/.append style={font=\normalsize},
  ipe stretch normal/.style={ipe node stretch=1},
  ipe stretch normal,
  ipe opacity 10/.style={opacity=0.1},
  ipe opacity 30/.style={opacity=0.3},
  ipe opacity 50/.style={opacity=0.5},
  ipe opacity 75/.style={opacity=0.75},
  ipe opacity opaque/.style={opacity=1},
  ipe opacity opaque,
]
\definecolor{red}{rgb}{1,0,0}
\definecolor{blue}{rgb}{0,0,1}
\definecolor{green}{rgb}{0,1,0}
\definecolor{yellow}{rgb}{1,1,0}
\definecolor{orange}{rgb}{1,0.647,0}
\definecolor{gold}{rgb}{1,0.843,0}
\definecolor{purple}{rgb}{0.627,0.125,0.941}
\definecolor{gray}{rgb}{0.745,0.745,0.745}
\definecolor{brown}{rgb}{0.647,0.165,0.165}
\definecolor{navy}{rgb}{0,0,0.502}
\definecolor{pink}{rgb}{1,0.753,0.796}
\definecolor{seagreen}{rgb}{0.18,0.545,0.341}
\definecolor{turquoise}{rgb}{0.251,0.878,0.816}
\definecolor{violet}{rgb}{0.933,0.51,0.933}
\definecolor{darkblue}{rgb}{0,0,0.545}
\definecolor{darkcyan}{rgb}{0,0.545,0.545}
\definecolor{darkgray}{rgb}{0.663,0.663,0.663}
\definecolor{darkgreen}{rgb}{0,0.392,0}
\definecolor{darkmagenta}{rgb}{0.545,0,0.545}
\definecolor{darkorange}{rgb}{1,0.549,0}
\definecolor{darkred}{rgb}{0.545,0,0}
\definecolor{lightblue}{rgb}{0.678,0.847,0.902}
\definecolor{lightcyan}{rgb}{0.878,1,1}
\definecolor{lightgray}{rgb}{0.827,0.827,0.827}
\definecolor{lightgreen}{rgb}{0.565,0.933,0.565}
\definecolor{lightyellow}{rgb}{1,1,0.878}
\definecolor{black}{rgb}{0,0,0}
\definecolor{white}{rgb}{1,1,1}
\begin{tikzpicture}[ipe stylesheet]
  \draw[shift={(128, 764)}, xscale=0.1875, yscale=0.875]
    (0, 0) rectangle (128, -64);
  \draw[shift={(108, 736)}, xscale=0.625, ->]
    (0, 0)
     -- (32, 0);
  \node[ipe node, rotate=90, anchor=center]
     at (100, 736) {Input text};
  \node[ipe node, rotate=90, anchor=center]
     at (140, 736) {Tokenizer};
  \begin{scope}[shift={(24, -13.5155)}]
    \node[ipe node, anchor=center]
       at (184, 692) {Encoding};
    \node[ipe node, anchor=center]
       at (184, 704) {Positional};
  \end{scope}
  \begin{scope}[shift={(-100.879, -216.312)}, xscale=1.2361, yscale=1.309]
    \draw
      (224, 700) circle[radius=8.9443];
    \begin{scope}[shift={(2, 0)}]
      \draw
        (216, 700)
         arc[start angle=180, end angle=360, x radius=3, y radius=-4];
      \draw
        (222, 700)
         arc[start angle=180, end angle=360, x radius=3, y radius=4];
    \end{scope}
  \end{scope}
  \node[ipe node, anchor=center, font=\large]
     at (176, 736) {$\oplus$};
  \draw[shift={(152, 736)}, xscale=0.625, ->]
    (0, 0)
     -- (32, 0);
  \draw[->]
    (176, 712)
     -- (176, 732);
  \draw[shift={(260, 700)}, rotate=90, xscale=0.5625, yscale=0.4375]
    (0, 0) rectangle (128, -64);
  \draw[shift={(180, 736)}, xscale=0.625, ->]
    (0, 0)
     -- (32, 0);
  \draw[shift={(308, 716)}, rotate=90, xscale=0.3125, yscale=0.375]
    (0, 0) rectangle (128, -64);
  \node[ipe node, rotate=90, anchor=center]
     at (320, 736) {Pooler};
  \draw[shift={(288, 736)}, xscale=0.625, ->]
    (0, 0)
     -- (32, 0);
  \node[ipe node, rotate=90, anchor=base, font=\small]
     at (301.96, 762.531) {$(\text{B}, \text{L}, 768)$};
  \node[ipe node, rotate=90, anchor=base, font=\small]
     at (389.96, 762.531) {$(\text{B}, 1, 192)$};
  \draw[shift={(376, 736)}, xscale=0.625, ->]
    (0, 0)
     -- (32, 0);
  \draw[shift={(396, 708)}, rotate=90, xscale=0.4375, yscale=0.375]
    (0, 0) rectangle (128, -64);
  \node[ipe node, rotate=90, anchor=center]
     at (408, 736) {Classifier};
  \node[ipe node, rotate=90, anchor=base, font=\small]
     at (433.96, 762.531) {$(\text{B}, 1, 921)$};
  \draw[shift={(420, 736)}, xscale=0.625, ->]
    (0, 0)
     -- (32, 0);
  \node[ipe node, rotate=90, anchor=center]
     at (489.93, 736) {Speaker logits};
  \draw[shift={(200, 764)}, xscale=0.1563, yscale=0.875]
    (0, 0) rectangle (128, -64);
  \node[ipe node, rotate=90, anchor=center]
     at (210, 736) {LayerNorm};
  \draw[shift={(220, 764)}, xscale=0.1563, yscale=0.875]
    (0, 0) rectangle (128, -64);
  \node[ipe node, rotate=90, anchor=center]
     at (230, 736) {Dropout};
  \draw[shift={(240, 736)}, xscale=0.625, ->]
    (0, 0)
     -- (32, 0);
  \node[ipe node, rotate=90, anchor=center, font=\small]
     at (268.3, 736) {(12 x)};
  \node[ipe node, rotate=90, anchor=center]
     at (280.3, 736) {BERT Layer};
  \draw[shift={(440, 700)}, rotate=90, xscale=0.5625, yscale=0.375]
    (0, 0) rectangle (128, -64);
  \node[ipe node, rotate=90, anchor=center]
     at (452, 736) {AAM-Softmax};
  \draw[shift={(464, 736)}, xscale=0.625, ->]
    (0, 0)
     -- (32, 0);
  \draw[shift={(332, 736)}, xscale=0.625, ->]
    (0, 0)
     -- (32, 0);
  \node[ipe node, rotate=90, anchor=center]
     at (364, 736) {Penultimate};
  \node[ipe node, rotate=90, anchor=base, font=\small]
     at (345.96, 762.531) {$(\text{B}, 1, 768)$};
  \draw[shift={(352, 708)}, rotate=90, xscale=0.4375, yscale=0.375]
    (0, 0) rectangle (128, -64);
  \draw[shift={(120, 664)}, rotate=90, xscale=1.0625, yscale=3.375, ipe dash dotted]
    (0, 0) rectangle (128, -64);
  \node[ipe node, anchor=west]
     at (128, 790.485) {\texttt{BertForSequenceClassification}};
  \draw[white]
    (496, 784)
     -- (496, 696);
\end{tikzpicture}\endgroup \renewcommand{\baselinestretch}{1.5}}}\\ \vspace{-0.16em}
    \subfloat[Enrollment]{\resizebox{0.95\linewidth}{!}{\begingroup
\renewcommand{\baselinestretch}{1} \endlinechar=-1 \tikzstyle{ipe stylesheet} = [
  ipe import,
  even odd rule,
  line join=round,
  line cap=butt,
  ipe pen normal/.style={line width=0.4},
  ipe pen heavier/.style={line width=0.8},
  ipe pen fat/.style={line width=1.2},
  ipe pen ultrafat/.style={line width=2},
  ipe pen normal,
  ipe mark normal/.style={ipe mark scale=3},
  ipe mark large/.style={ipe mark scale=5},
  ipe mark small/.style={ipe mark scale=2},
  ipe mark tiny/.style={ipe mark scale=1.1},
  ipe mark normal,
  /pgf/arrow keys/.cd,
  ipe arrow normal/.style={scale=7},
  ipe arrow large/.style={scale=10},
  ipe arrow small/.style={scale=5},
  ipe arrow tiny/.style={scale=3},
  ipe arrow normal,
  /tikz/.cd,
  ipe arrows, 
  <->/.tip = ipe normal,
  ipe dash normal/.style={dash pattern=},
  ipe dash dotted/.style={dash pattern=on 1bp off 3bp},
  ipe dash dashed/.style={dash pattern=on 4bp off 4bp},
  ipe dash dash dotted/.style={dash pattern=on 4bp off 2bp on 1bp off 2bp},
  ipe dash dash dot dotted/.style={dash pattern=on 4bp off 2bp on 1bp off 2bp on 1bp off 2bp},
  ipe dash normal,
  ipe node/.append style={font=\normalsize},
  ipe stretch normal/.style={ipe node stretch=1},
  ipe stretch normal,
  ipe opacity 10/.style={opacity=0.1},
  ipe opacity 30/.style={opacity=0.3},
  ipe opacity 50/.style={opacity=0.5},
  ipe opacity 75/.style={opacity=0.75},
  ipe opacity opaque/.style={opacity=1},
  ipe opacity opaque,
]
\definecolor{red}{rgb}{1,0,0}
\definecolor{blue}{rgb}{0,0,1}
\definecolor{green}{rgb}{0,1,0}
\definecolor{yellow}{rgb}{1,1,0}
\definecolor{orange}{rgb}{1,0.647,0}
\definecolor{gold}{rgb}{1,0.843,0}
\definecolor{purple}{rgb}{0.627,0.125,0.941}
\definecolor{gray}{rgb}{0.745,0.745,0.745}
\definecolor{brown}{rgb}{0.647,0.165,0.165}
\definecolor{navy}{rgb}{0,0,0.502}
\definecolor{pink}{rgb}{1,0.753,0.796}
\definecolor{seagreen}{rgb}{0.18,0.545,0.341}
\definecolor{turquoise}{rgb}{0.251,0.878,0.816}
\definecolor{violet}{rgb}{0.933,0.51,0.933}
\definecolor{darkblue}{rgb}{0,0,0.545}
\definecolor{darkcyan}{rgb}{0,0.545,0.545}
\definecolor{darkgray}{rgb}{0.663,0.663,0.663}
\definecolor{darkgreen}{rgb}{0,0.392,0}
\definecolor{darkmagenta}{rgb}{0.545,0,0.545}
\definecolor{darkorange}{rgb}{1,0.549,0}
\definecolor{darkred}{rgb}{0.545,0,0}
\definecolor{lightblue}{rgb}{0.678,0.847,0.902}
\definecolor{lightcyan}{rgb}{0.878,1,1}
\definecolor{lightgray}{rgb}{0.827,0.827,0.827}
\definecolor{lightgreen}{rgb}{0.565,0.933,0.565}
\definecolor{lightyellow}{rgb}{1,1,0.878}
\definecolor{black}{rgb}{0,0,0}
\definecolor{white}{rgb}{1,1,1}
\begin{tikzpicture}[ipe stylesheet]
  \draw[shift={(136, 764)}, xscale=0.1875, yscale=0.875]
    (0, 0) rectangle (128, -64);
  \draw[shift={(116, 736)}, xscale=0.625, ->]
    (0, 0)
     -- (32, 0);
  \node[ipe node, rotate=90, anchor=center]
     at (100, 736) {Enrolment utt.};
  \node[ipe node, rotate=90, anchor=center]
     at (148, 736) {Tokenizer};
  \begin{scope}[shift={(32, -13.5155)}]
    \node[ipe node, anchor=center]
       at (184, 692) {Encoding};
    \node[ipe node, anchor=center]
       at (184, 704) {Positional};
  \end{scope}
  \begin{scope}[shift={(-92.879, -216.312)}, xscale=1.2361, yscale=1.309]
    \draw
      (224, 700) circle[radius=8.9443];
    \begin{scope}[shift={(2, 0)}]
      \draw
        (216, 700)
         arc[start angle=180, end angle=360, x radius=3, y radius=-4];
      \draw
        (222, 700)
         arc[start angle=180, end angle=360, x radius=3, y radius=4];
    \end{scope}
  \end{scope}
  \node[ipe node, anchor=center, font=\large]
     at (184, 736) {$\oplus$};
  \draw[shift={(160, 736)}, xscale=0.625, ->]
    (0, 0)
     -- (32, 0);
  \draw[->]
    (184, 712)
     -- (184, 732);
  \draw[shift={(268, 700)}, rotate=90, xscale=0.5625, yscale=0.4375]
    (0, 0) rectangle (128, -64);
  \draw[shift={(188, 736)}, xscale=0.625, ->]
    (0, 0)
     -- (32, 0);
  \draw[shift={(316, 716)}, rotate=90, xscale=0.3125, yscale=0.375]
    (0, 0) rectangle (128, -64);
  \node[ipe node, rotate=90, anchor=center]
     at (328, 736) {Pooler};
  \draw[shift={(296, 736)}, xscale=0.625, ->]
    (0, 0)
     -- (32, 0);
  \node[ipe node, rotate=90, anchor=base, font=\small]
     at (309.96, 778.531) {$(\text{N}_\text{enroll\_utt}, \text{L}, 768)$};
  \node[ipe node, rotate=90, anchor=base, font=\small]
     at (353.96, 778.531) {$(\text{N}_\text{enroll\_utt}, 1, 768)$};
  \draw[shift={(340, 736)}, xscale=0.625, ->]
    (0, 0)
     -- (32, 0);
  \draw[shift={(360, 708)}, rotate=90, xscale=0.4375, yscale=0.375]
    (0, 0) rectangle (128, -64);
  \node[ipe node, rotate=90, anchor=center]
     at (372, 736) {Penultimate};
  \node[ipe node, rotate=90, anchor=base, font=\small]
     at (397.96, 778.531) {$(\text{N}_\text{enroll\_utt}, 1, 192)$};
  \draw[shift={(384, 736)}, xscale=0.625, ->]
    (0, 0)
     -- (32, 0);
  \draw[shift={(208, 764)}, xscale=0.1563, yscale=0.875]
    (0, 0) rectangle (128, -64);
  \node[ipe node, rotate=90, anchor=center]
     at (218, 736) {LayerNorm};
  \draw[shift={(228, 764)}, xscale=0.1563, yscale=0.875]
    (0, 0) rectangle (128, -64);
  \node[ipe node, rotate=90, anchor=center]
     at (238, 736) {Dropout};
  \draw[shift={(248, 736)}, xscale=0.625, ->]
    (0, 0)
     -- (32, 0);
  \node[ipe node, rotate=90, anchor=center, font=\small]
     at (276.3, 736) {(12 x)};
  \node[ipe node, rotate=90, anchor=center]
     at (288.3, 736) {BERT Layer};
  \node[ipe node, rotate=90, anchor=center]
     at (112, 736) {Input text};
  \draw[shift={(404, 708)}, rotate=90, xscale=0.4375, yscale=0.375]
    (0, 0) rectangle (128, -64);
  \node[ipe node, rotate=90, anchor=center]
     at (440, 736) {Average};
  \draw[shift={(452, 736)}, xscale=0.625, ->]
    (0, 0)
     -- (32, 0);
  \node[ipe node, rotate=90, anchor=base, font=\small]
     at (465.96, 762.531) {$(1, 1, 192)$};
  \node[ipe node, rotate=90, anchor=center]
     at (490, 736) {Enrolment embedding};
  \draw[shift={(428, 708)}, rotate=90, xscale=0.4375, yscale=0.375]
    (0, 0) rectangle (128, -64);
  \node[ipe node, rotate=90, anchor=center]
     at (416, 736) {L2Norm};
  \draw[white]
    (496, 776)
     -- (496, 688);
  \node[ipe node]
     at (96, 804) {(for each enrollment speaker)};
  \node[ipe node]
     at (96, 788) {(compute utt. embeddings independently)};
  \node[ipe node, rotate=90, anchor=center]
     at (478, 736) {(spk. level)};
\end{tikzpicture}\endgroup \renewcommand{\baselinestretch}{1.5}}}\\ \vspace{-0.17em}
    \subfloat[Trial]{\resizebox{0.95\linewidth}{!}{\begingroup
\renewcommand{\baselinestretch}{1} \endlinechar=-1 \tikzstyle{ipe stylesheet} = [
  ipe import,
  even odd rule,
  line join=round,
  line cap=butt,
  ipe pen normal/.style={line width=0.4},
  ipe pen heavier/.style={line width=0.8},
  ipe pen fat/.style={line width=1.2},
  ipe pen ultrafat/.style={line width=2},
  ipe pen normal,
  ipe mark normal/.style={ipe mark scale=3},
  ipe mark large/.style={ipe mark scale=5},
  ipe mark small/.style={ipe mark scale=2},
  ipe mark tiny/.style={ipe mark scale=1.1},
  ipe mark normal,
  /pgf/arrow keys/.cd,
  ipe arrow normal/.style={scale=7},
  ipe arrow large/.style={scale=10},
  ipe arrow small/.style={scale=5},
  ipe arrow tiny/.style={scale=3},
  ipe arrow normal,
  /tikz/.cd,
  ipe arrows, 
  <->/.tip = ipe normal,
  ipe dash normal/.style={dash pattern=},
  ipe dash dotted/.style={dash pattern=on 1bp off 3bp},
  ipe dash dashed/.style={dash pattern=on 4bp off 4bp},
  ipe dash dash dotted/.style={dash pattern=on 4bp off 2bp on 1bp off 2bp},
  ipe dash dash dot dotted/.style={dash pattern=on 4bp off 2bp on 1bp off 2bp on 1bp off 2bp},
  ipe dash normal,
  ipe node/.append style={font=\normalsize},
  ipe stretch normal/.style={ipe node stretch=1},
  ipe stretch normal,
  ipe opacity 10/.style={opacity=0.1},
  ipe opacity 30/.style={opacity=0.3},
  ipe opacity 50/.style={opacity=0.5},
  ipe opacity 75/.style={opacity=0.75},
  ipe opacity opaque/.style={opacity=1},
  ipe opacity opaque,
]
\definecolor{red}{rgb}{1,0,0}
\definecolor{blue}{rgb}{0,0,1}
\definecolor{green}{rgb}{0,1,0}
\definecolor{yellow}{rgb}{1,1,0}
\definecolor{orange}{rgb}{1,0.647,0}
\definecolor{gold}{rgb}{1,0.843,0}
\definecolor{purple}{rgb}{0.627,0.125,0.941}
\definecolor{gray}{rgb}{0.745,0.745,0.745}
\definecolor{brown}{rgb}{0.647,0.165,0.165}
\definecolor{navy}{rgb}{0,0,0.502}
\definecolor{pink}{rgb}{1,0.753,0.796}
\definecolor{seagreen}{rgb}{0.18,0.545,0.341}
\definecolor{turquoise}{rgb}{0.251,0.878,0.816}
\definecolor{violet}{rgb}{0.933,0.51,0.933}
\definecolor{darkblue}{rgb}{0,0,0.545}
\definecolor{darkcyan}{rgb}{0,0.545,0.545}
\definecolor{darkgray}{rgb}{0.663,0.663,0.663}
\definecolor{darkgreen}{rgb}{0,0.392,0}
\definecolor{darkmagenta}{rgb}{0.545,0,0.545}
\definecolor{darkorange}{rgb}{1,0.549,0}
\definecolor{darkred}{rgb}{0.545,0,0}
\definecolor{lightblue}{rgb}{0.678,0.847,0.902}
\definecolor{lightcyan}{rgb}{0.878,1,1}
\definecolor{lightgray}{rgb}{0.827,0.827,0.827}
\definecolor{lightgreen}{rgb}{0.565,0.933,0.565}
\definecolor{lightyellow}{rgb}{1,1,0.878}
\definecolor{black}{rgb}{0,0,0}
\definecolor{white}{rgb}{1,1,1}
\begin{tikzpicture}[ipe stylesheet]
  \draw[shift={(136, 764)}, xscale=0.1875, yscale=0.875]
    (0, 0) rectangle (128, -64);
  \draw[shift={(116, 736)}, xscale=0.625, ->]
    (0, 0)
     -- (32, 0);
  \node[ipe node, rotate=90, anchor=center]
     at (100, 736) {Trial utt.};
  \node[ipe node, rotate=90, anchor=center]
     at (148, 736) {Tokenizer};
  \begin{scope}[shift={(32, -13.5155)}]
    \node[ipe node, anchor=center]
       at (184, 692) {Encoding};
    \node[ipe node, anchor=center]
       at (184, 704) {Positional};
  \end{scope}
  \begin{scope}[shift={(-92.879, -216.312)}, xscale=1.2361, yscale=1.309]
    \draw
      (224, 700) circle[radius=8.9443];
    \begin{scope}[shift={(2, 0)}]
      \draw
        (216, 700)
         arc[start angle=180, end angle=360, x radius=3, y radius=-4];
      \draw
        (222, 700)
         arc[start angle=180, end angle=360, x radius=3, y radius=4];
    \end{scope}
  \end{scope}
  \node[ipe node, anchor=center, font=\large]
     at (184, 736) {$\oplus$};
  \draw[shift={(160, 736)}, xscale=0.625, ->]
    (0, 0)
     -- (32, 0);
  \draw[->]
    (184, 712)
     -- (184, 732);
  \draw[shift={(268, 700)}, rotate=90, xscale=0.5625, yscale=0.4375]
    (0, 0) rectangle (128, -64);
  \draw[shift={(188, 736)}, xscale=0.625, ->]
    (0, 0)
     -- (32, 0);
  \draw[shift={(316, 716)}, rotate=90, xscale=0.3125, yscale=0.375]
    (0, 0) rectangle (128, -64);
  \node[ipe node, rotate=90, anchor=center]
     at (328, 736) {Pooler};
  \draw[shift={(296, 736)}, xscale=0.625, ->]
    (0, 0)
     -- (32, 0);
  \node[ipe node, rotate=90, anchor=base, font=\small]
     at (309.96, 762.531) {$(1, \text{L}, 768)$};
  \node[ipe node, rotate=90, anchor=base, font=\small]
     at (353.96, 762.531) {$(1, 1, 768)$};
  \draw[shift={(340, 736)}, xscale=0.625, ->]
    (0, 0)
     -- (32, 0);
  \draw[shift={(208, 764)}, xscale=0.1563, yscale=0.875]
    (0, 0) rectangle (128, -64);
  \node[ipe node, rotate=90, anchor=center]
     at (218, 736) {LayerNorm};
  \draw[shift={(228, 764)}, xscale=0.1563, yscale=0.875]
    (0, 0) rectangle (128, -64);
  \node[ipe node, rotate=90, anchor=center]
     at (238, 736) {Dropout};
  \draw[shift={(248, 736)}, xscale=0.625, ->]
    (0, 0)
     -- (32, 0);
  \node[ipe node, rotate=90, anchor=center, font=\small]
     at (276.3, 736) {(12 x)};
  \node[ipe node, rotate=90, anchor=center]
     at (288.3, 736) {BERT Layer};
  \node[ipe node, rotate=90, anchor=center]
     at (112, 736) {Input text};
  \node[ipe node]
     at (96, 788) {(for each trial utterance)};
  \node[ipe node, anchor=base, font=\small]
     at (375.717, 688.288) {$(\text{N}_\text{enroll\_spk}, 1, 192)$};
  \draw[->]
    (376, 696)
     -- (376, 700)
     -- (392, 700)
     -- (392, 720)
     -- (404, 720);
  \node[ipe node, anchor=base]
     at (376, 672) {Enrollment embeddings (spk. level)};
  \draw[shift={(360, 708)}, rotate=90, xscale=0.4375, yscale=0.375]
    (0, 0) rectangle (128, -64);
  \node[ipe node, rotate=90, anchor=center]
     at (372, 736) {Penultimate};
  \node[ipe node, rotate=90, anchor=base, font=\small]
     at (397.96, 762.531) {$(1, 1, 192)$};
  \draw[shift={(384, 736)}, xscale=0.625, ->]
    (0, 0)
     -- (32, 0);
  \draw[white]
    (496, 776)
     -- (496, 688);
  \draw
    (404, 744) rectangle (460, 712);
  \node[ipe node, anchor=center]
     at (432, 734) {Cosine};
  \node[ipe node, anchor=center]
     at (432, 722) {Similarity};
\end{tikzpicture}\endgroup \renewcommand{\baselinestretch}{1.5}}} \vspace{-0.17em}
    \vspace{-0.5em}
    \caption{Training, enrollment, and trial phases of our system.}
    \label{fig:text_based_attack}
    \vspace{-1.5em}
\end{figure}

We initialize our network with the checkpoint available online\footnote{\url{https://huggingface.co/google-bert/bert-base-uncased}}, and we use normal initialization for new layers. Then, we fine-tune on \texttt{train-clean-360}, using Speechbrain's implementation \cite{ravanelli_open-source_2024} of \ac{AAM} softmax loss \cite{xiang_margin_2019}, and setting the involved hyperparameters as summarized in Table~\ref{tab:hparams}. The training-validation split is performed such that both subsets contain utterances from each speaker and each session (called \texttt{spk-diverse-sess} in the \ac{VPC} 2024 codebase). During training, we track the loss and the classification accuracy on the hold-out validation split. We reserve \texttt{libri-dev} and \texttt{libri-test} for evaluation purposes. After six epochs of training, the validation accuracy converges to 54\%, while the validation loss starts to increase. We interpret this as the model starting to overfit, and to avoid that, we use the checkpoint after six training epochs for evaluations in the upcoming sections.

During our experiments, we noticed that the L2 norms of the utterance-level embeddings differ, even though the network encodes information exclusively via the direction of the embeddings. Therefore, unlike \ac{VPC} 2024 $\text{ASV}_\textrm{eval}^\textrm{anon}$, our text-based attack normalizes the utterance-level enrollment vectors to unit length (see Fig.~\ref{fig:text_based_attack}b). This is to facilitate equal contribution of each utterance towards the speaker-level enrollment vector. The source code to reproduce our work will be released at \footnote{\url{https://doi.org/10.5281/zenodo.15526086}}.

\section{Results and discussion}

\begin{figure}[!t]
    \centering
    \resizebox{\linewidth}{!}{\begingroup
\renewcommand{\baselinestretch}{1} \endlinechar=-1 \input{Figures/results_text_based_interspeech.pgf}\endgroup \renewcommand{\baselinestretch}{1.5}}
    \vspace{-2em}
    \caption{Score distributions of our attack on \texttt{libri-dev}.}
    \label{fig:results}
    \vspace{-1.5em}
\end{figure}

\begin{figure}[!t]
    \centering
    \resizebox{0.87\linewidth}{!}{\begingroup
\renewcommand{\baselinestretch}{1} \endlinechar=-1 \input{Figures/results_spider_interspeech.pgf}\endgroup \renewcommand{\baselinestretch}{1.5}}
    \vspace{-1.5em}
    \caption{Radar plots comparing our text-based attack to $\text{ASV}_\textrm{eval}^\textrm{anon}$ on \texttt{libri-test} dataset. Spokes indicate enrollment speaker IDs; circular axes show corresponding speaker \acp{EER}}
    \label{fig:results_spider}
    \vspace{-1.25em}
\end{figure}

Fig.~\ref{fig:results} shows the score distributions of our text-based attack; please refer to Fig.~\ref{fig:vpc_scores_breakdown} for comparison and interpretation. On average, our attack achieves an \ac{EER} of 33.68\% for female speakers and 36.30\% for male speakers, performing only slightly worse than $\text{ASV}_\textrm{eval}^\textrm{anon}$ despite the limited available information. 

Turning to speaker-level performance, speakers \texttt{1673} and \texttt{652}, the ones that were identified in Sec.~\ref{sec:status_quo}, achieved \acp{EER} of 1.60\% and 16.81\%, respectively, indicating that their anonymity was in fact compromised by the text-based attack. In Fig.~\ref{fig:results_spider}, we include a radar plot to visualize how our attack compares to $\text{ASV}_\textrm{eval}^\textrm{anon}$ on \texttt{libri-test} subset. There, our attack achieved an \ac{EER} of 4.10\% and 11.86\% for speakers \texttt{3570} and \texttt{2830}.

Also, we found that L2-normalizing utterance-level embeddings for enrollment slightly improved the performance of our system, reducing the mean \ac{EER} by 0.39 and 0.32 percentage points for female and male speakers, respectively. We think the effects of normalization are worth exploring for $\text{ASV}_\textrm{eval}^\textrm{anon}$.

\vspace{0.5\baselineskip}
\noindent \textbf{Explainability analysis}
\vspace{0.5\baselineskip}

\begin{figure}[!t]
    \setlength{\fboxsep}{1pt}
    \centering
    \subfloat[Speaker \texttt{1673}]{\resizebox{\linewidth}{!}{\includegraphics[]{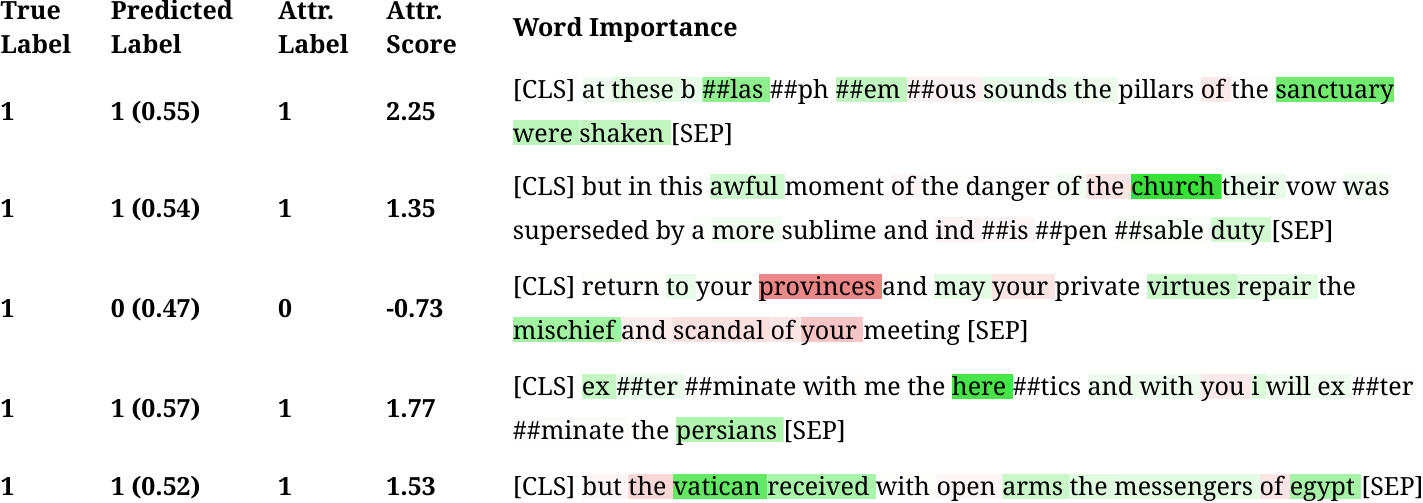}}}\\
    \subfloat[Speaker \texttt{652}]{\resizebox{\linewidth}{!}{\includegraphics[]{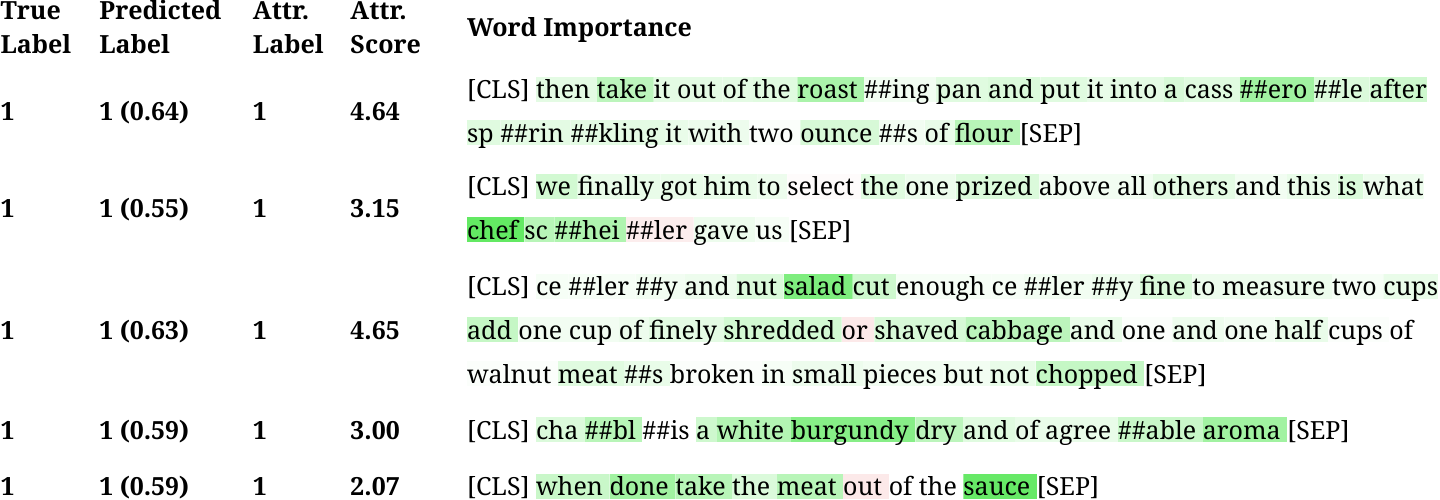}}}\\
    \subfloat[Speaker \texttt{7976}]{\resizebox{\linewidth}{!}{\includegraphics[]{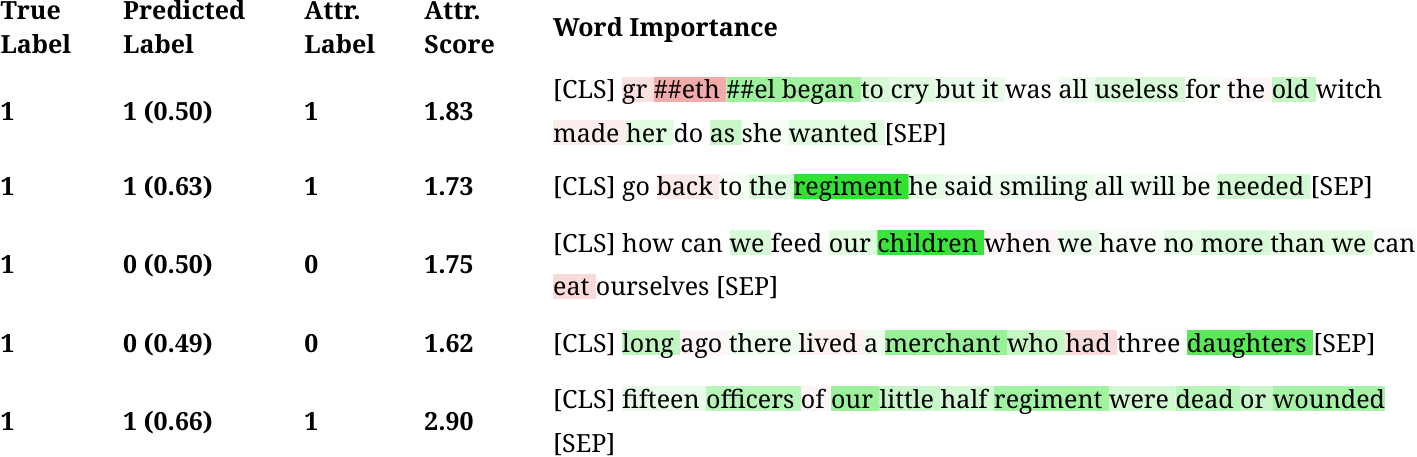}}}\\
    \caption{Explainability study results. Tokens that contribute positively to a model decision are highlighted in \colorbox{green}{green}, and \colorbox{red}{red} stands for negative contributions. The intensity of the highlight signifies the strength. Attribution score is the sum of all word importance scores, used as a measure of confidence. The characters '\#\#' occur when a word is represented by multiple tokens to show that surrounding tokens are part of the same word.}
    \label{fig:results_transformers_interpret}
    \vspace{-1em}
\end{figure}

\noindent We use \texttt{transformers-interpret} \cite{pierse_transformers-interpret_2021, zhu_using_2022}, a Python package, to get insight into our model decisions. Specifically, we apply Layer Integrated Gradients \cite{sundararajan_axiomatic_2017} to the cosine similarities in reference to \ac{EER} thresholds (see Fig.~\ref{fig:results}). We selected five random trial utterances for three speakers, and used the corresponding speaker-level embedding for enrollment. The resulting attribution scores and word importance scores are visualized in Fig.~\ref{fig:results_transformers_interpret} for two successfully attacked speakers (\texttt{1673} and \texttt{652}) and one failed attack (\texttt{7976}). Overall, the successful attacks are linked to semantically similar keywords, such as religious terms for speaker \texttt{1673} (church, Vatican, heretics, ...), and culinary terms for speaker \texttt{652} (meat, salad, casserole, ...). In contrast, for speaker \texttt{7976}, while some utterances with military terms (regiment, officer, ...) were recognized, many utterances did not have these, so the attack failed (\ac{EER}: 50.11\%).

\vspace{0.5\baselineskip}
\noindent \textbf{Discussion}
\vspace{0.5\baselineskip}


\noindent Existence of speakers for which the text-based attack is successful suggests that the dataset has exploitable intra-speaker linguistic content similarity. This finding is particularly significant as our model achieves comparable verification performance to $\text{ASV}_\textrm{eval}^\textrm{anon}$ despite only having access to the text in \texttt{libri-test}, while $\text{ASV}_\textrm{eval}^\textrm{anon}$ using the anonymized speech.

Beyond global \ac{EER} values, our text-based attack and $\text{ASV}_\textrm{eval}^\textrm{anon}$ show different behaviors at the speaker level. For certain speakers, e.g., \texttt{6829} in \texttt{libri-test}, we have observed that the text-based attack failed, but $\text{ASV}_\textrm{eval}^\textrm{anon}$ has been successful. The other case is, e.g., \texttt{1284} or \texttt{2830} in \texttt{libri-test}, where we see that the text-based attack was successful, but (at least some) anonymization systems still managed to attain decent \acp{EER}. Besides, the \acp{EER} vary for successful and failed attacks, such as our text-based attack achieves 1.60\% EER for speaker \texttt{1673} but the semi-informed attacks on B3, B4 and B5 attain 11.03\%, 12.54\% and 15.65\%, respectively.

Several factors can explain these variations. First, the success of $\text{ASV}_\textrm{eval}^\textrm{anon}$ in cases where our text-based attack failed may be attributed to the additional information available in speech signals, such as speaking rate and fundamental frequency.

Conversely, our text-based attack's occasional superior performance is likely due to its more sophisticated linguistic understanding, enabled by using a pre-trained BERT model. Notably, attempts to train our attack without pre-trained BERT weights failed to converge. Similarly, our initial experiments using classic NLP methods such as TD-IDF and CountVectorizer also failed, even after preprocessing to remove overly common words. In contrast, $\text{ASV}_\textrm{eval}^\textrm{anon}$, powered by ECAPA-TDNN, is unlikely to learn and utilize semantic similarity of the thematic words to the extent of the text-based attack. Second factor is the possible changes in linguistic content caused by the anonymization process, e.g., speech degradation leading to increased word error rates. We think investigating the effects of anonymization on linguistic content would constitute an interesting follow-up study. Nevertheless, our findings highlight an important vulnerability: speakers can be recognized through their corresponding text, and the risk of attackers exploiting this grows as their architectures become more advanced.


\section{Conclusion}

In this work, we explored the speaker-level behavior of $\text{ASV}_\textrm{eval}^\textrm{anon}$ on speaker anonymization systems. In our analysis, we identified that reporting global \acp{EER}, which is a common practice in evaluating \ac{ASV} systems for speaker verification, can obfuscate the shortcomings of speaker anonymization systems by overestimating their effectiveness. To tackle this issue, we proposed reporting average \ac{EER} after clipping speaker-level \acp{EER} exceeding 50\%. Furthermore, we identified some speakers in the evaluation datasets, whose anonymity were repeatedly compromised. To investigate if \ac{VPC} speakers can be identified solely by their linguistic content, we repurposed BERT as an \ac{ASV} model. Our system achieves \acp{EER} less than 20\% for 4 / 29 enrolled speakers in \texttt{libri-dev} subset and 6 / 29 enrolled speakers in \texttt{libri-test}, showing that the linguistic content similarity in the utterances of these speakers is sufficient to verify their identities. Our explainability analysis suggests model decisions are influenced by thematically similar keywords, such as culinary or religious terms. Further work is needed to develop clean datasets for \ac{VPC} attack training and evaluations and to quantify how much attacks in the literature exploit this.

\section{Acknowledgements}
This work is partially supported by the German Ministry of Research, Technology and Space (BMFTR) under grant agreement No. 16IS24072F (COMFORT).

\printbibliography[heading=bibnumbered]

\end{document}